\documentclass[epj,referee]{svjour}
\usepackage{graphics}
\usepackage{graphics}
\usepackage{amssymb}
\usepackage{amsmath}
\usepackage{graphicx}
\usepackage{dcolumn}
\usepackage{bm}

\begin{document}

\title{BOUNDS ON BIPARTITIELY SHARED ENTANGLEMENT REDUCED FROM SUPERPOSED
TRIPARTITE QUANTUM STATES}
\author{Chang-shui Yu\thanks{\emph{Email address:}quaninformation@sina.com
or ycs@dlut.edu.cn} \and X. X. Yi \and He-shan Song }
\institute{Department of Physics, Dalian University of Technology, Dalian 116024, P. R.
China}
\date{Received: date / Revised version: date}



\abstract{ For a tripartite pure state superposed by two individual
states, the bipartitely shared entanglement can always be achieved
by local measurements of the third party. Consider the different
aims of the third party, i.e. maximizing or minimizing the
bipartitely shared entanglement, we find bounds on both the possible
bipartitely shared entanglement of the superposition state in terms
of the corresponding entanglement of the two states being
superposed. In particular, by choosing the concurrence as bipartite
entanglement measure, we obtain calculable bounds for tripartite
$(2\otimes 2\otimes n)$ -dimensional cases. \PACS{
      {}{03.65.Ud }   \and
      {}{03.67.Mn}\and
      {}{03.65.Ta}
     } } %end of abstract
\maketitle
\section{\protect\bigskip Introduction}

How the entanglement of a state depends on the entanglement of individual
states in the superposition is the key to well understand the nature of
entanglement. It also plays an important role in understanding the
entanglement of a mixed state. This question was first addressed by Linden,
Popescu and Smolin in Ref. [1], who employed von Neumann entropy of the
reduced matrix as entanglement measure (ER) and presented an upper bound on
ER of the superposed states in terms of those of the states being
superposed. The von Neumann entropy of $\left\vert \Psi \right\rangle $ was
defined [2] by%
\begin{equation}
E(\left\vert \Psi \right\rangle )=S(\rho _{x})=-\text{Tr}\rho _{x}\log
_{2}\rho _{x},
\end{equation}%
where $\rho _{x}=$Tr$_{x}\left\vert \Psi \right\rangle \left\langle \Psi
\right\vert $ denotes the reduced density matrix of A or B. Recently this
problem becomes active and has been extensively studied. Yu et al have
studied the concurrence of superpositions [3] and presented an upper bound
and a lower bound on the concurrence of superpositions. Ou et al [4] give an
upper bound on the negativity of superpositions. Niset and Cerf [5]
reconsidered the concurrence of superpositions and gave lower and upper
bounds simpler than ours. Jitesh R. Bhatt et al [6] addressed the problem by
considering two superposed coherent states. Davalvanti et al [7] and Song et
al [8] have addressed the entanglement of superpositions for multipartite
quantum states by employing different entanglement measures. Most recently,
Gour [9] reconsidered the question in Ref. [1] and presented tighter upper
and lower bounds.

In this paper, we consider the bounds on entanglement induced by
superposition in a new and very interesting case. As we know, one of the
important methods to producing bipartite entanglement is the reduction of a
multipartite entangled state to an entangled state over fewer parties (e.g.
bipartite) via measurements [10]. A natural and very interesting question is
that if a tripartite quantum pure state superposed by two individual states
is considered, how the entanglement shared by two parties with the third
party's assistance is influenced by the superposition of the two individual
states (Here it implies that von Neumann entropy is employed as bipartite
entanglement measure). Suppose that tripartite pure states are shared by
Alice, Bob and Charlie and we concern bipartite entanglement shared by Alice
and Bob. There are two different possibilities considering Charlie's
different aims. One is to minimize the entanglement between Alice and Bob
described by bipartite entanglement of their reduced density matrix (say
\emph{bipartite entanglement} in the following for simplification), the
other is to maximize the entanglement with assistance of Charlie described
by the entanglement of assistance (EoA) [10,11].

We will show how bipartitely shared entanglement on the corresponding
entanglement of individual states in the superpositions\footnote{%
In other words, we consider the entanglement of reduced density matrix of
the superposed tripartite quantum pure states. One should note that it is
essentially different from directly considering the entanglement of a given
ensemble. The reasons are as follows. 1) The reduced density matrix of two
superposed tripartite pure states is obviously different from the classical
sum of two corresponding mixed states. 2) For a tripartite pure state, the
entanglement shared by two parties can be adjusted by the assistance of the
third party; For a give ensemble (or mixed state), its entanglement only
denotes the minimal average entanglement of all pure-state realizations, it
is no sense in saying the maximal entanglement of a given ensemble. 3) Since
EOA (in particular COA) characterizes tripartite entanglement instead of
bipartite entanglement [12], what we are considering is tripartite
entanglement instead of bipartite entanglement. Our system should include 3
subsystems instead of only bipartite mixed states}. Upper bounds on
bipartite entanglement and EoA are derived in terms of the entanglement of
the states being superposed. The corresponding lower bounds can be naturally
obtained analogously to Ref. [5,8] and the generalization to the case of the
superposition of more than two terms is straightforward. Both are briefly
stated here. What is more, because bipartite entanglement and EoA are
inconvenient to calculate and in particular, EoA is not an entanglement
monotone [12], we also consider that Alice, Bob and Charlie share tripartite
$\left( 2\otimes 2\otimes n\right) $ -dimensional pure states and find the
bounds on the entanglement bipartitely shared by Alice and Bob by employing
concurrence as bipartite entanglement measure. In this case, the minimal and
the maximal bipartitely shared entanglement are given, respectively, by
Wootters' concurrence [13] and concurrence of assistance (CoA) [11,14,15]
which is shown to be a tripartite entanglement monotone [11].

The paper is organized as follows. In Sec. II, by utilizing von Neumann
entropy as bipartite entanglement measure, we derive upper bounds on
bipartitely shared entanglement of superposed tripartite quantum states. In
Sec. III, calculable bounds on bipartitely shared entanglement are found for
$(2\otimes 2\otimes n)$-dimensional quantum states in terms of bipartite
concurrence. The conclusion is drawn finally.

\section{Bounds in terms of von Neumann Entropy}

Given a tripartite quantum pure state $\left\vert \Upsilon \right\rangle
_{ABC}$, the entanglement of reduced density matrix can be given by
\begin{equation}
E\left( \rho _{AB}\right) =\min \sum\limits_{i}p_{i}E\left( \left\vert
\Lambda \right\rangle _{AB}\right)
\end{equation}%
and EoA is defined by [11]
\begin{equation}
E_{a}\left( \left\vert \Upsilon \right\rangle _{ABC}\right) =\max
\sum\limits_{i}p_{i}E\left( \left\vert \Lambda \right\rangle _{AB}\right) ,
\end{equation}%
where the minimum and the maximum are taken over all possible decompositions
of
\begin{equation}
\rho _{AB}=\text{Tr}_{C}\left( \left\vert \Upsilon \right\rangle
_{ABC}\left\langle \Upsilon \right\vert \right)
=\sum\limits_{i}p_{i}\left\vert \Lambda \right\rangle _{AB}\left\langle
\Lambda \right\vert .
\end{equation}%
Note that the subscript $ABC$ throughout of the paper denotes Alice, Bob and
Charlie, respectively.

\textbf{Theorem 1.} \textit{Suppose that }$\left\vert \Phi \right\rangle
_{ABC}$\textit{\ and }$\left\vert \Psi \right\rangle _{ABC}$\textit{\ are
two normalized tripartite pure states, and }$\left\vert \Gamma \right\rangle
_{ABC}=\alpha \left\vert \Phi \right\rangle _{ABC}+\beta \left\vert \Psi
\right\rangle _{ABC}$\textit{\ be the superposed state with }$\left\vert
\alpha \right\vert ^{2}+\left\vert \beta \right\vert ^{2}=1$\textit{. Let }$%
\varrho _{AB}=$Tr$_{C}\left( \left\vert \Gamma \right\rangle
_{ABC}\left\langle \Gamma \right\vert \right) $\textit{, }$\varrho
_{_{1}AB}= $Tr$_{C}\left( \left\vert \Phi \right\rangle
_{ABC}\left\langle \Phi \right\vert \right) $\textit{\ and }$\varrho
_{_{2}AB}=$ Tr$_{C}\left(
\left\vert \Psi \right\rangle _{ABC}\left\langle \Psi \right\vert \right) $%
\textit{, then }%
\begin{gather}
\left\vert \left\vert \Gamma \right\rangle _{ABC}\right\vert ^{2}E\left(
\varrho _{AB}\right) \leq \left\vert \alpha \right\vert ^{2}\left[ E\left(
\varrho _{_{1}AB}\right) +E_{a}\left( \left\vert \Phi \right\rangle
_{ABC}\right) \right]   \notag \\
+\left\vert \beta \right\vert ^{2}\left[ E_{a}\left( \left\vert \Psi
\right\rangle _{ABC}\right) +E\left( \varrho _{_{2}AB}\right) \right]
+4\left\vert \alpha \beta \right\vert ,
\end{gather}%
\textit{and}%
\begin{eqnarray}
&&\left\vert \left\vert \Gamma \right\rangle _{ABC}\right\vert
^{2}E_{a}\left( \left\vert \Gamma \right\rangle _{ABC}\right)   \notag \\
&\leq &2\left[ \left\vert \alpha \right\vert ^{2}E_{a}\left( \left\vert \Phi
\right\rangle _{ABC}\right) +\left\vert \beta \right\vert ^{2}E_{a}\left(
\left\vert \Psi \right\rangle _{ABC}\right) +2\left\vert \alpha \beta
\right\vert \right] ,
\end{eqnarray}%
\textit{where }$\left\vert \left\vert \Gamma \right\rangle _{ABC}\right\vert
$\textit{\ denotes }$l_{2}$\textit{\ norm of }$\left\vert \Gamma
\right\rangle _{ABC}$\textit{.}

\textbf{Proof. }Based on\textbf{\ }HJW theorem [16], any ensemble that
represents $\varrho _{AB}$ can be achieved by the local Positive Operator
Value Measurements (POVM) [17] on Charlie's system. Let $M_{j}=I_{AB}\otimes
N_{j}$, $\sum_{j}N_{j}N_{j}^{\dag }=1$ be the POVM operators on Charlie's
party written in terms of Kraus operators [14], then the corresponding
ensemble can be given by $\{p_{ij},\left\vert \Upsilon _{ij}\right\rangle
_{AB}\}$ where
\begin{equation}
\sqrt{p_{ij}}\left\vert \Upsilon _{ij}\right\rangle _{AB}=\frac{\left\langle
i\right\vert M_{j}\left\vert \Gamma \right\rangle _{ABC}}{\left\vert
\left\vert \Gamma \right\rangle _{ABC}\right\vert }
\end{equation}%
with
\begin{equation}
p_{ij}=\frac{\left\vert \left\langle i\right\vert M_{j}\left\vert \Gamma
\right\rangle _{ABC}\right\vert ^{2}}{\left\vert \left\vert \Gamma
\right\rangle _{ABC}\right\vert ^{2}}
\end{equation}%
and $\left\vert i\right\rangle $ being the computational basis of Charlie's
system. The average entanglement is $\sum\limits_{ij}p_{ij}E\left(
\left\vert \Upsilon _{ij}\right\rangle _{AB}\right) \ $by which we have
(Note that we have omitted the subscripts for simplification.)
\begin{gather}
\sum\limits_{ij}p_{ij}E\left( \left\vert \Upsilon _{ij}\right\rangle \right)
=\sum\limits_{ij}p_{ij}  \notag \\
\times E\left( \alpha \sqrt{\frac{q_{1ij}}{p_{ij}\left\vert \left\vert
\Gamma \right\rangle \right\vert ^{2}}}\left\vert \Phi _{ij}^{\prime
}\right\rangle +\beta \sqrt{\frac{q_{2ij}}{p_{ij}\left\vert \left\vert
\Gamma \right\rangle \right\vert ^{2}}}\left\vert \Psi _{ij}^{\prime
}\right\rangle \right)   \notag \\
\leq 2\sum\limits_{ij}p_{ij}\left[ \left\vert \alpha \sqrt{\frac{q_{1ij}}{%
p_{ij}\left\vert \left\vert \Gamma \right\rangle \right\vert ^{2}}}%
\right\vert ^{2}E\left( \left\vert \Phi _{ij}^{\prime }\right\rangle \right)
\right.   \notag \\
+\left\vert \beta \sqrt{\frac{q_{2ij}}{p_{ij}\left\vert \left\vert \Gamma
\right\rangle \right\vert ^{2}}}\right\vert ^{2}E\left( \left\vert \Psi
_{ij}^{\prime }\right\rangle \right)   \notag \\
\left. +2\left\vert \alpha \sqrt{\frac{q_{1ij}}{p_{ij}\left\vert \left\vert
\Gamma \right\rangle \right\vert ^{2}}}\right\vert \left\vert \beta \sqrt{%
\frac{q_{2ij}}{p_{ij}\left\vert \left\vert \Gamma \right\rangle \right\vert
^{2}}}\right\vert \right]   \notag \\
=\frac{2}{\left\vert \left\vert \Gamma \right\rangle \right\vert ^{2}}%
\sum\limits_{ij}\left[ \left\vert \alpha \right\vert ^{2}q_{1ij}E\left(
\left\vert \Phi _{ij}^{\prime }\right\rangle \right) \right.   \notag \\
\left. +\left\vert \beta \right\vert ^{2}q_{2ij}E\left( \left\vert
\Psi _{ij}^{\prime }\right\rangle \right) +2\left\vert \alpha \beta
\right\vert \sqrt{q_{1ij}q_{2ij}}\right] ,
\end{gather}%
where
\begin{equation*}
\left\vert \Phi ^{\prime }\right\rangle =\frac{\left\langle i\right\vert
M_{j}\left\vert \Phi \right\rangle }{\sqrt{q_{1ij}}},\left\vert \Psi
^{\prime }\right\rangle =\frac{\left\langle i\right\vert M_{j}\left\vert
\Psi \right\rangle }{\sqrt{q_{2ij}}},
\end{equation*}%
\begin{equation*}
q_{1ij}=\left\vert \left\langle i\right\vert M_{j}\left\vert \Phi
\right\rangle \right\vert ,q_{2ij}=\left\vert \left\langle i\right\vert
M_{j}\left\vert \Psi \right\rangle \right\vert .
\end{equation*}%
In addition, the first inequality in eq. (9) follows from the inequality
[11]
\begin{equation}
h(x)\leq 2\sqrt{x(1-x)},
\end{equation}%
with
\begin{equation}
h(x)=-x\log _{2}x-(1-x)\log _{2}(1-x),
\end{equation}%
and the original bound on entanglement of superposed states given in Ref.
[1] by replacing $\alpha $ and $\beta $ in Ref. [1] with $\alpha \sqrt{\frac{%
q_{1ij}}{p_{ij}\left\vert \left\vert \Gamma \right\rangle \right\vert ^{2}}}$
and $\beta \sqrt{\frac{q_{2ij}}{p_{ij}\left\vert \left\vert \Gamma
\right\rangle \right\vert ^{2}}}$, respectively. Now, if we suppose $\left\{
q_{1ij},\left\vert \Phi _{ij}^{\prime }\right\rangle \right\} $ is the
optimal decomposition of $\varrho _{1AB}$ in the sense of minimal average
entanglement, it is obvious that
\begin{eqnarray}
&&\left\vert \left\vert \Gamma \right\rangle \right\vert ^{2}E\left( \varrho
_{AB}\right)   \notag \\
&\leq &2\left( \left\vert \alpha \right\vert ^{2}E\left( \varrho
_{_{1}AB}\right) +\left\vert \beta \right\vert ^{2}E_{a}\left( \left\vert
\Psi \right\rangle _{ABC}\right) +2\left\vert \alpha \beta \right\vert
\right) ,
\end{eqnarray}%
where we apply the definitions of $E\left( \varrho _{AB}\right) $ and $%
E_{a}\left( \left\vert \Psi \right\rangle _{ABC}\right) $ and the
Cauchy-Schwarz inequality
\begin{equation}
\sum_{ij}\sqrt{q_{1ij}q_{2ij}}\leq 1.
\end{equation}%
Analogously, let $\left\{ q_{1ij},\left\vert \Phi _{ij}^{\prime
}\right\rangle \right\} $ minimize the entanglement of $\varrho _{2AB}$,
then
\begin{equation}
\left\vert \left\vert \Gamma \right\rangle \right\vert ^{2}E\left( \varrho
_{AB}\right) \leq 2\left( \left\vert \alpha \right\vert ^{2}E_{a}\left(
\left\vert \Phi \right\rangle _{ABC}\right) +\left\vert \beta \right\vert
^{2}E\left( \varrho _{_{2}AB}\right) +2\left\vert \alpha \beta \right\vert
\right) .
\end{equation}%
Therefore, inequalities (12) and (14) can also be written in a more
symmetric form as inequality (5). If $\left\{ p_{ij},\left\vert \Upsilon
_{ij}\right\rangle _{AB}\right\} $ is supposed to be the optimal
decomposition in the sense of maximal average entanglement, one will obtain%
\begin{eqnarray}
&&\left\vert \left\vert \Gamma \right\rangle \right\vert
^{2}E_{a}\left( \left\vert \Gamma \right\rangle \right) \notag \\
&\leq &2\left( \left\vert \alpha \right\vert ^{2}E_{a}\left(
\left\vert \Phi \right\rangle _{ABC}\right) +\left\vert \beta
\right\vert ^{2}E_{a}\left( \left\vert \Psi \right\rangle
_{ABC}\right) +2\left\vert \alpha \beta \right\vert \right) .
\end{eqnarray}%
The proof is completed.\hfill {}$\Box $

\section{ Bounds in terms of concurrence}

We have employed von Neumann Entropy as bipartite entanglement measure to
investigate the minimal and the maximal entanglement shared by Alice and Bob
with the assistance of Charlie, provided that they share a tripartite
quantum pure state. As we know, in general EoA and bipartite entanglement of
mixed states are difficult to calculate. In particular, EoA is not an
entanglement monotone. In this sense, EoA of superposition seems not to be a
good candidate for the investigation, even though one can not worry about
the EoA in the bound of $E\left( \varrho _{AB}\right) $ where EoA can be
only considered as a function. On the contrary, it has been shown that $%
\left( 2\otimes 2\right) $-dimensional concurrence and CoA of the $(2\otimes
2\otimes n)$- dimensional pure states are explicitly given and CoA is a good
entanglement measure. Next we restrict tripartite quantum states to $%
(2\otimes 2\otimes n)$- dimensional case. CoA of a tripartite pure state $%
\left\vert \Upsilon ^{\prime }\right\rangle _{ABC}$ can be defined [10,15] by%
\begin{equation}
C_{a}\left( \left\vert \Upsilon ^{\prime }\right\rangle _{ABC}\right) =\max
\sum\limits_{i}p_{i}^{\prime }C\left( \left\vert \Lambda _{i}^{\prime
}\right\rangle _{AB}\right) =F\left( \rho _{AB}^{\prime },\tilde{\rho}%
_{AB}^{\prime }\right) ,
\end{equation}%
where%
\begin{equation}
\rho _{AB}^{\prime }=Tr_{C}\left\vert \Upsilon ^{\prime }\right\rangle
_{ABC}\left\langle \Upsilon ^{\prime }\right\vert ,
\end{equation}%
\begin{equation}
\tilde{\rho}_{AB}^{\prime }=\left( \sigma _{y}\otimes \sigma _{y}\right)
\rho _{AB}^{\prime \ast }\left( \sigma _{y}\otimes \sigma _{y}\right)
\end{equation}%
and
\begin{equation}
F\left( \sigma ^{\prime },\rho ^{\prime }\right) =Tr\sqrt{\sqrt{\sigma
^{\prime }}\rho ^{\prime }\sqrt{\sigma ^{\prime }}}.
\end{equation}%
The concurrence of $\rho _{AB}^{\prime }$ is defined by [13]%
\begin{eqnarray}
C\left( \rho _{AB}^{\prime }\right)  &=&\min \sum\limits_{i}p_{i}C\left(
\left\vert \Lambda _{i}^{\prime }\right\rangle _{AB}\right)  \\
&=&\max \{0,\lambda _{1}-\sum_{i>1}\lambda _{i}\},
\end{eqnarray}%
with $\lambda _{i}$ being the square roots of eigenvalues of $\tilde{\rho}%
_{AB}^{\prime }\rho _{AB}^{\prime }$ in decreasing order. The extremums in
eq. (16) and eq. (20) are taken over all possible decompositions of $\rho
_{AB}^{\prime }$.

\textbf{Theorem 2:} \textit{Let }$\left\vert \Phi ^{\prime }\right\rangle
_{ABC}$\textit{\ and }$\left\vert \Psi ^{\prime }\right\rangle _{ABC}$%
\textit{\ be two }$(2\otimes 2\otimes n)$\textit{- dimensional pure states.
The superposition state is given by }$\left\vert \Gamma ^{\prime
}\right\rangle _{ABC}=\alpha \left\vert \Phi ^{\prime }\right\rangle
_{ABC}+\beta \left\vert \Psi ^{\prime }\right\rangle _{ABC}$\textit{, }$%
\left\vert \alpha \right\vert ^{2}+\left\vert \beta \right\vert ^{2}=1.$%
\textit{\ Let }$\varrho _{AB}^{\prime }=$Tr$_{C}\left( \left\vert \Gamma
^{\prime }\right\rangle _{ABC}\left\langle \Gamma ^{\prime }\right\vert
\right) $\textit{, }$\varrho _{_{1}AB}^{\prime }=$Tr$_{C}\left( \left\vert
\Phi ^{\prime }\right\rangle _{ABC}\left\langle \Phi ^{\prime }\right\vert
\right) $\textit{\ and }$\varrho _{_{2}AB}^{\prime }=$Tr$_{C}\left(
\left\vert \Psi ^{\prime }\right\rangle _{ABC}\left\langle \Psi ^{\prime
}\right\vert \right) $\textit{, then }%
\begin{gather}
\left\vert \left\vert \Gamma ^{\prime }\right\rangle _{ABC}\right\vert
^{2}C\left( \varrho _{AB}^{\prime }\right) \leq \frac{\left\vert \alpha
\right\vert ^{2}}{2}\left[ C\left( \varrho _{_{1}AB}^{\prime }\right)
+C_{a}\left( \left\vert \Phi ^{\prime }\right\rangle _{ABC}\right) \right]
\notag \\
+\frac{\left\vert \beta \right\vert ^{2}}{2}\left[ C_{a}\left( \left\vert
\Psi ^{\prime }\right\rangle _{ABC}\right) +C\left( \varrho
_{_{2}AB}^{\prime }\right) \right] +2\left\vert \alpha \beta \right\vert ,
\end{gather}%
\textit{and}%
\begin{eqnarray}
&&\left\vert \left\vert \Gamma ^{\prime }\right\rangle _{ABC}\right\vert
^{2}C_{a}\left( \left\vert \Gamma ^{\prime }\right\rangle _{ABC}\right)  \notag\\
&\leq &\left\vert \alpha \right\vert ^{2}C_{a}\left( \left\vert \Phi
^{\prime }\right\rangle _{ABC}\right) +\left\vert \beta \right\vert
^{2}C_{a}\left( \left\vert \Psi ^{\prime }\right\rangle _{ABC}\right)
+2\left\vert \alpha \beta \right\vert ,
\end{eqnarray}%
\textit{where }$\left\vert \left\vert \Gamma ^{\prime }\right\rangle
_{ABC}\right\vert $\textit{\ denotes }$l_{2}$\textit{\ norm of }$\left\vert
\Gamma ^{\prime }\right\rangle _{ABC}$\textit{.}

\textbf{Proof. }Let\textbf{\ }$\{p_{ij}^{\prime },\left\vert \Upsilon
_{ij}^{\prime }\right\rangle _{AB}\}$ be any\ ensemble that represents $%
\varrho _{AB}^{\prime }$ corresponding to a POVM operation $M_{j}$ on
Charlie's system analogous to that in \textbf{Theorem 1}. The average
concurrence is given by $\sum\limits_{ij}p_{ij}^{\prime }C\left( \left\vert
\Upsilon _{ij}^{\prime }\right\rangle _{AB}\right) $ where
\begin{equation}
\sqrt{p_{ij}^{\prime }}\left\vert \Upsilon _{ij}^{\prime }\right\rangle
_{AB}=\frac{\left\langle i\right\vert M_{j}\left\vert \Gamma ^{\prime
}\right\rangle _{ABC}}{\left\vert \left\vert \Gamma ^{\prime }\right\rangle
_{ABC}\right\vert }
\end{equation}%
with
\begin{equation}
p_{ij}^{\prime }=\frac{\left\vert \left\langle i\right\vert M_{j}\left\vert
\Gamma ^{\prime }\right\rangle _{ABC}\right\vert ^{2}}{\left\vert \left\vert
\Gamma ^{\prime }\right\rangle _{ABC}\right\vert ^{2}}
\end{equation}%
\begin{equation}
\left\vert \Phi _{ij}^{\prime \prime }\right\rangle =\frac{\left\langle
i\right\vert M_{j}\left\vert \Phi ^{\prime }\right\rangle }{\sqrt{%
q_{1ij}^{\prime }}},\left\vert \Psi _{ij}^{\prime \prime }\right\rangle =%
\frac{\left\langle i\right\vert M_{j}\left\vert \Psi ^{\prime }\right\rangle
}{\sqrt{q_{2ij}^{\prime }}},
\end{equation}%
\begin{equation}
q_{1ij}^{\prime }=\left\vert \left\langle i\right\vert M_{j}\left\vert \Phi
^{\prime }\right\rangle \right\vert ,q_{2ij}^{\prime }=\left\vert
\left\langle i\right\vert M_{j}\left\vert \Psi ^{\prime }\right\rangle
\right\vert .
\end{equation}%
Note that we have omitted the subscripts for simplification again. Thus,

\begin{gather}
\sum\limits_{ij}p_{ij}^{\prime }C\left( \left\vert \Upsilon _{ij}^{\prime
}\right\rangle \right)  \notag \\
=\sum\limits_{ij}p_{ij}^{\prime }C\left( \alpha \sqrt{\frac{q_{1ij}^{\prime }%
}{p_{ij}^{\prime }\left\vert \left\vert \Gamma ^{\prime }\right\rangle
\right\vert ^{2}}}\left\vert \Phi _{ij}^{\prime \prime }\right\rangle +\beta
\sqrt{\frac{q_{2ij}^{\prime }}{p_{ij}^{\prime }\left\vert \left\vert \Gamma
^{\prime }\right\rangle \right\vert ^{2}}}\left\vert \Psi _{ij}^{\prime
\prime }\right\rangle \right)  \notag \\
\leq \sum\limits_{ij}p_{ij}^{\prime }\left[ \left\vert \alpha \sqrt{\frac{%
q_{1ij}^{\prime }}{p_{ij}\left\vert \left\vert \Gamma ^{\prime
}\right\rangle \right\vert ^{2}}}\right\vert ^{2}\left\vert \left\langle
\Phi _{ij}^{\prime \prime \ast }\right\vert \sigma _{y}\otimes \sigma
_{y}\left\vert \Phi _{ij}^{\prime \prime }\right\rangle \right\vert \right.
\notag \\
+\left\vert \beta \sqrt{\frac{q_{2ij}^{\prime }}{p_{ij}^{\prime }\left\vert
\left\vert \Gamma ^{\prime }\right\rangle \right\vert ^{2}}}\right\vert
^{2}\left\vert \left\langle \Psi _{ij}^{\prime \prime \ast }\right\vert
\sigma _{y}\otimes \sigma _{y}\left\vert \Phi _{ij}^{\prime \prime
}\right\rangle \right\vert  \notag \\
+\left. 2\left\vert \alpha \beta \sqrt{\frac{q_{1ij}^{\prime }}{%
p_{ij}^{\prime }\left\vert \left\vert \Gamma ^{\prime }\right\rangle
\right\vert ^{2}}}\sqrt{\frac{q_{2ij}^{\prime }}{p_{ij}^{\prime }\left\vert
\left\vert \Gamma ^{\prime }\right\rangle \right\vert ^{2}}}\left\langle
\Psi _{ij}^{\prime \prime \ast }\right\vert \sigma _{y}\otimes \sigma
_{y}\left\vert \Phi _{ij}^{\prime \prime }\right\rangle \right\vert \right]
\notag \\
\leq \frac{1}{\left\vert \left\vert \Gamma ^{\prime }\right\rangle
\right\vert ^{2}}\sum\limits_{ij}\left[ \left\vert \alpha \right\vert
^{2}q_{1ij}^{\prime }C\left( \left\vert \Phi _{ij}^{\prime \prime
}\right\rangle \right) \right. \left. +\left\vert \beta \right\vert
^{2}q_{2ij}^{\prime }C\left( \left\vert \Psi _{ij}^{\prime \prime
}\right\rangle \right) \right]  \notag \\
+2\left\vert \alpha \beta \right\vert .
\end{gather}%
Here the first inequality follows from the definition of concurrence for
pure state and the inequality $\sum\limits_{k}\left\vert z_{k}\right\vert
\geq \left\vert \sum\limits_{k}z_{k}\right\vert $. The second inequality is
derived from the inequality (13) and
\begin{equation}
\left\vert \left\langle \Psi _{ij}^{\prime \prime \ast }\right\vert
\sigma _{y}\otimes \sigma _{y}\left\vert \Phi _{ij}^{\prime \prime
}\right\rangle \right\vert \leq \left\vert \left\vert \Psi
_{ij}^{\prime \prime }\right\rangle \right\vert \cdot \left\vert
\left\vert \Phi _{ij}^{\prime \prime }\right\rangle \right\vert =1.
\end{equation}%
If we suppose $\left\{ q_{1ij}^{\prime \prime },\left\vert \Phi
_{ij}^{\prime \prime }\right\rangle \right\} $ is the optimal decomposition
of $\varrho _{1AB}^{\prime }$ in the sense of minimal average concurrence,
we have
\begin{equation}
\left\vert \left\vert \Gamma ^{\prime }\right\rangle \right\vert ^{2}C\left(
\varrho _{AB}^{\prime }\right) \leq \left\vert \alpha \right\vert
^{2}C\left( \varrho _{_{1}AB}^{\prime }\right) +\left\vert \beta \right\vert
^{2}C_{a}\left( \left\vert \Psi ^{\prime }\right\rangle _{ABC}\right)
+2\left\vert \alpha \beta \right\vert .
\end{equation}%
Let $\left\{ q_{2ij}^{\prime \prime },\left\vert \Psi _{ij}^{\prime
\prime }\right\rangle \right\} $ minimize the average concurrence of
$\varrho _{2AB}^{\prime }$, then
\begin{equation}
\left\vert \left\vert \Gamma ^{\prime }\right\rangle \right\vert ^{2}C\left(
\varrho _{AB}^{\prime }\right) \leq \left\vert \alpha \right\vert
^{2}C_{a}\left( \left\vert \Phi ^{\prime }\right\rangle _{ABC}\right)
+\left\vert \beta \right\vert ^{2}C\left( \varrho _{_{2}AB}^{\prime }\right)
+2\left\vert \alpha \beta \right\vert .
\end{equation}%
Inequalities (30) and (31) can be rewritten in the more symetric form as
inequality (22). If $\left\{ p_{ij}^{\prime },\left\vert \Upsilon
_{ij}^{\prime }\right\rangle \right\} $ is supposed to be the optimal
decomposition maximizing the average concurrence, one will get%
\begin{equation}
\left\vert \left\vert \Gamma ^{\prime }\right\rangle \right\vert
^{2}C_{a}\left( \left\vert \Gamma ^{\prime }\right\rangle \right) \leq
\left\vert \alpha \right\vert ^{2}C_{a}\left( \left\vert \Phi ^{\prime
}\right\rangle _{ABC}\right) +\left\vert \beta \right\vert ^{2}C_{a}\left(
\left\vert \Psi ^{\prime }\right\rangle _{ABC}\right) +2\left\vert \alpha
\beta \right\vert .
\end{equation}%
The proof is completed.\hfill {}$\Box $

\begin{figure}[tbp]
\centering
\includegraphics[width=8cm]{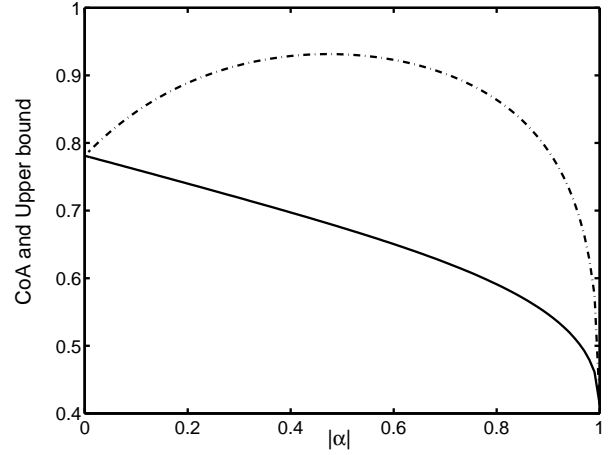}% Here is how to import EPS art
\caption{(dimensionless) The upper bound on CoA and the CoA of the random
generated state $\left\vert\protect\gamma\right\rangle_{ABC}$ vs the
absolute value $\left\vert\protect\alpha\right\vert$. The dash-dot line
corresponds to the upper bound of CoA, and the solid line is the CoA of $%
\left\vert\protect\gamma\right\rangle_{ABC}$.}
\end{figure}
\begin{figure}[tbp]
\centering
\includegraphics[width=8cm]{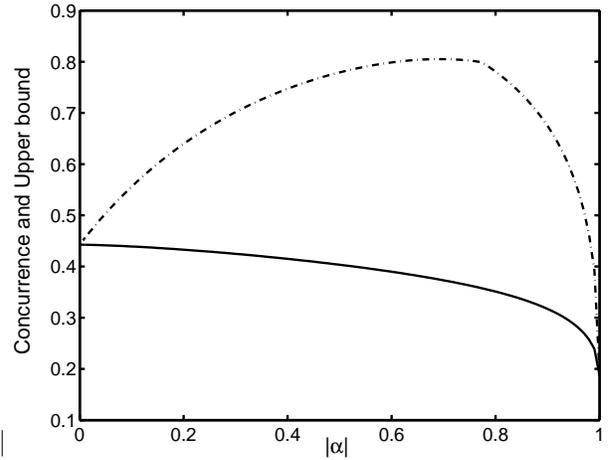}% Here is how to import EPS art
\caption{(dimensionless) Concurrence of $Tr_C\left\vert\protect\gamma%
\right\rangle_{ABC}\left\langle\protect\gamma\right\vert$ and its upper
bound vs the absolute value $\left\vert\protect\alpha\right\vert$. The
dash-dot line corresponds to the upper bound of concurrence, and the solid
line is the concurrence of $Tr_C\left\vert\protect\gamma\right\rangle_{ABC}%
\left\langle\protect\gamma\right\vert$.}
\end{figure}

As an application, by \textbf{Matlab 6.5} we generate two random $(2\otimes
2\otimes 4)$- dimensional pure states
\begin{eqnarray}
\left\vert \phi \right\rangle _{ABC} &=&[0.4061,0.1119,0.1321,0.4155,  \notag
\\
&&0.2188,0.3618,0.0422,0.3351,  \notag \\
&&0.2407,0.1541,0.1120,0.0759,  \notag \\
&&0.2656,0.2659,0.2019,0.2402]^{T},
\end{eqnarray}%
and

\begin{eqnarray}
\left\vert \psi \right\rangle _{ABC} &=&[0.3868,0.0250,0.4408,0.0716,  \notag
\\
&&0.1171,0.1588,0.1093,0.0930,  \notag \\
&&0.0581,0.2613,0.1253,0.0290,  \notag \\
&&0.2439,0.4571,0.3642,0.3189]^{T}.
\end{eqnarray}%
The relation between CoA of the superposed state $\left\vert \gamma
\right\rangle _{ABC}=\alpha \left\vert \phi \right\rangle _{ABC}+\beta
\left\vert \psi \right\rangle _{ABC}$, $\left\vert \alpha \right\vert
^{2}+\left\vert \beta \right\vert ^{2}=1$, and its upper bound with
different $\left\vert \alpha \right\vert $ and the relation between
bipartite concurrence and its upper bound are shown in Fig. 1 and Fig. 2,
respectively. In fact, we have random chosen $10^{6}$ pairs of $(2\otimes
2\otimes 4)$- dimensional $\left\vert \phi \right\rangle _{ABC}$ and $%
\left\vert \psi \right\rangle _{ABC}$, all the numerical results show that
\textbf{Theorem 2 } can provide good upper bounds. From Fig. 2 one can find
that the upper bound is as large as twice the actual value of concurrence.
however, the ratio of the upper bound and the actual concurrence is between
1 and 2. The minimal value corresponds to the superposition including a
W-type state [17] with the probability trending to 1 and the maximal one
corresponds to the superposition of two GHZ-type [17] state with equal
probabilities. In fact, the state-dependent tightness exists in all the
relevant works [1,3-9]. The \emph{tighter} bound still needs further efforts.

\section{Discussion and Conclusion}

Before the end, we would like to taking eq. (5) in \textbf{Theorem 1} as an
example (the others are analogous.) to briefly discuss the lower bounds and
the generalization to the case of the superposition of more than two terms.
At first, we briefly state the introduction of lower bound. The state $%
\left\vert \Gamma \right\rangle _{ABC}=\alpha \left\vert \Phi \right\rangle
_{ABC}+\beta \left\vert \Psi \right\rangle _{ABC}$\textit{\ }can always be
rewritten by
\begin{equation}
\left\vert \Phi \right\rangle _{ABC}=\frac{\left\vert \left\vert \Gamma
\right\rangle _{ABC}\right\vert }{\alpha }\frac{\left\vert \Gamma
\right\rangle _{ABC}}{\left\vert \left\vert \Gamma \right\rangle
_{ABC}\right\vert }-\frac{\beta }{\alpha }\left\vert \Psi \right\rangle
_{ABC}.
\end{equation}%
Apply eq. (5) in \textbf{Theorem 1} to eq. (35), one can obtain the
analogous bound on $E(\varrho _{_{1}AB})$. The upper bound includes $E\left(
\varrho _{AB}\right) $ and $E_{a}\left( \left\vert \Gamma \right\rangle
_{ABC}\right) $ which is further limited by eq. (6). On the contrary, a
lower bound on $E\left( \varrho _{AB}\right) $ can be derived from the upper
bound. An analogous lower bound based on the analogous expression to eq.
(35) for $\left\vert \Psi \right\rangle _{ABC}$ can also be obtained. The
minimal lower bound serves as a potential lower bound. Our theorems can be
straightforwardly generalized to the case of superposition of more than two
terms. Without loss of generality, suppose the superposition of three terms
as $\left\vert \Pi \right\rangle _{ABC}=a\left\vert \Phi \right\rangle
_{ABC}+b\left\vert \Psi \right\rangle _{ABC}+c\left\vert \Theta
\right\rangle _{ABC}$ which can be rewritten by
\begin{equation}
\left\vert \Pi \right\rangle _{ABC}=\sqrt{\left\vert a\right\vert
^{2}+\left\vert b\right\vert ^{2}}\left\vert \Gamma _{p}\right\rangle
_{ABC}+c\left\vert \Theta \right\rangle _{ABC}
\end{equation}%
with%
\begin{equation}
\left\vert \Gamma _{p}\right\rangle _{ABC}=\frac{a\left\vert \Phi
\right\rangle _{ABC}+b\left\vert \Psi \right\rangle _{ABC}}{\sqrt{\left\vert
a\right\vert ^{2}+\left\vert b\right\vert ^{2}}},
\end{equation}%
then we can apply eq. (5) to eq. (36) and obtain the upper bound on $E\left(
\sigma _{AB}\right) $ in terms of $E\left( \sigma _{AB1}\right) $, $E\left(
\sigma _{AB2}\right) $, $E_{a}\left( \sigma _{AB1}\right) $ and $E_{a}\left(
\sigma _{AB2}\right) $ where $\sigma _{AB}=$Tr$_{C}\left( \left\vert \Pi
\right\rangle _{ABC}\left\langle \Pi \right\vert \right) $, $\sigma _{AB1}=$%
Tr$_{C}\left( \left\vert \Gamma _{p}\right\rangle _{ABC}\left\langle \Gamma
_{p}\right\vert \right) $ and $\sigma _{AB2}=$Tr$_{C}\left( \left\vert
\Theta \right\rangle _{ABC}\left\langle \Theta \right\vert \right) $. Employ
eq. (5) and eq. (6) again on eq. (37), one can obtain upper bounds on $%
E\left( \sigma _{AB1}\right) $ and $E_{a}\left( \sigma _{AB1}\right) $,
respectively, which will lead to the final upper bound on $E\left( \sigma
_{AB}\right) $ in terms of the bipartitely shared entanglement of the three
superposed quantum states.

In summary, for a tripartite quantum pure state shared by Alice, Bob and
Charlie and superposed by two individual states, we have presented upper
bounds on the entanglement shared by Alice and Bob with assistance of
Charlie in terms of his different aims. We consider the bounds by employing
von Neumann entropy and concurrence respectively. The latter provides
calculable bounds for $(2\otimes 2\otimes n)$-dimensional quantum states. In
particular, although CoA maximizes the concurrence shared by Alice and Bob,
CoA is a tripartite entanglement monotone instead of bipartite entanglement
monotone [12]. It is worthy of being noted that the lower bounds can be
naturally obtained similarly to Ref. [5,8], which is only a simple algebra
and briefly stated here. What is more, the generalization of the presented
bounds to the case where there are more than two terms in the superposition
is straightforward. We believe the \emph{tighter} bounds on the entanglement
of superpositions is still important and interesting for further work.

\section{Acknowledgement}

Yu and Song thank the support by the National Natural Science
Foundation of China (NNSFC), under Grant No. 10747112, No. 10575017
and No.60703100. Yi thanks the support by NNSFC, under Grant No.
60578014.

\end{document}